\documentclass[aps,prb,preprint]{revtex4-2}
\pdfoutput=1
\usepackage{graphicx}
\usepackage{braket}
\usepackage{amsmath}
\usepackage{bm}
\usepackage{hyperref}
\newcommand{\ext}{\text{ex}}

\begin{document}

\title{Efficient homogenization of multicomponent metamaterials:
  chiral effects}

\author{W. Luis Mochán}
\email{mochan@fis.unam.mx}
\affiliation{Instituto de Ciencias F\'isicas, Universidad Nacional
  Aut\'onoma de M\'exico, Avenida Universidad s/n, 62210 Cuernavaca,
  Morelos, M\'exico}
\author{Guillermo P. Ortiz}
\affiliation{Departamento de Física, Facultad de Ciencias Exactas,
  Naturales y Agrimensura, Universidad Nacional del Nordeste, Corrientes, Argentina}
\begin{abstract}
We extend an efficient homogenization procedure based on a Haydock
representation of the microscopic wave operator for the calculation of
the macroscopic dielectric response of a periodic composite to the
case of an arbitrary number of components of arbitrary
composition. As a test, we apply our numerical procedure to the calculation of
the optical properties of a Bouligand structure, made of a large
number of anisotropic layers stacked on top of each other and
progresively rotated. This system consitutes a photonic crystal with
circularly polarized electromagnetic normal modes, naturally ocurring
in the cuticle of several arthropods, and which has a gap for one
helicity, which corresponds to the observation of circularly
polarized strong metallic like reflections. Our numerical procedure is
validated through its good agreement with the analytical solution for
this simple chiral system.
\end{abstract}
\maketitle
\section{Introduction}\label{s:introduction}
The importance of the optical properties of inhomogeneous artificial media such as
metamaterials and photonic
crystals made up of alternating particles
of diverse ordinary materials has been well established. They have led
to novel miniature optical
devices \cite{zheludevMetamaterialsMetadevices2012}. Metamaterials may
display exotic optical properties that are absent from ordinary materials
such as a negative index of
refraction \cite{ThreedimensionalOpticalMetamaterial}. These properties
may me employed to design systems such as invisibility
cloaks \cite{leonhardtBroadbandInvisibilityNonEuclidean2009} based on
metamaterials with permittivities and permeabilities that mimic the
properties of curved space \cite{ulfleonhardtOpticalConformalMapping2006}.
By manipulating the variation of
the phase of a lightwave along a surface on which high index of
refraction particles of appropiate shape, size and orientation are
layed down, planar {\em metalenses}
\cite{AdvantagesMetalensesDiffractive} can be fabricated. They may be
designed to focus light with a high numerical aperture, to elliminate
chromatic aberration \cite{BroadbandAchromaticMetalens}, they may be
tuned through mechanical stretching or controlled phase changes changes
\cite{shalaginovReconfigurableAlldielectricMetalens2021}, and they may be sensitive to the
polarization of light. For example, in
Ref. \cite{wangCompactMultifociMetalens2023} a submillimeter
metasurface capable of focusing light of different wavelengths at
pre-established points along a single focal plane was developed, leading to
ultra-compact spectrometers.

One approach to the calculation of the optical properties of these
systems is through the homogenization of its dielectric properties;
the inhomogeneous system is replaced by a homogeneous material with
effective properties related to its composition and geometry
\cite{silveirinhaMetamaterialHomogenizationApproach2007a}. The most
simple and widely used homogenization method is that of
Maxwell-Garnett \cite{MaxwellGarnett(1904)}, based on the quasi-static
interaction among small polarizable spherical particles within a host. Many other
homogenization schemes have since been devised with a more general
applicability. For example, in
Ref. \cite{reyes-avendanoPhotonicCrystalsMetamaterials2011} a
formulation of the bianisotropic macroscopic response of a
metamaterial with an arbitrary composition and geometry was derived in
the long-wavelength regime.
A very general approach to the calculation of the
macroscopic dielectric response of inhomogeneous materials, with some
applications to liquids and disordered systems, roughnes at surfaces,
periodic crystals and the surface local field effect was developed in Refs.
\cite{mochanElectromagneticResponseSystems1985,mochanElectromagneticResponseSystems1985a}. The
formal theory therein has been further developed in the non-retarded
limit \cite{mochanEfficientHomogenizationProcedure2010}, and led to the
development of a very efficient numerical procedure, inspired on
Haydock's recursive procedure \cite{Haydock(1980)} for the calculation
of the projected Green's functions of quantum systems. That procedure
is applicable to binary metamaterials made of arbitrary materials, dielectric
or conducting, dispersive or not, dissipative or transparent, as it
factors out the dynamical response of the components, which are
encoded by a single number $u$, the possibly complex and frequency
dependent {\em spectral variable},
from the geometry of the system which is characterized by a real {\em
  characteristic function} $B(\bm r)$ of the position {$\bm r$}. The longitudinal
component of the characteristic function is interpreted as a Hermitian
operator, analogous to a Hamiltonian, and an iterative procedure
yields its representation as a symmetric tridiagonal matrix whose
elements are the Haydock coefficients. The macroscopic response is
then given by a continued fraction in terms of the Haydock
coefficients and the spectral variable. This procedure may be applied
in one, two or three dimensions
\cite{mochanEfficientHomogenizationProcedure2010,cortesOpticalPropertiesNanostructured2010},
and it has been used to design and optimize thin devices to manipulate the
polarization of light
\cite{mendozaBirefringentNanostructuredComposite2012,mendozaTailoredOpticalPolarization2016}.
Furthermore, the formalism allows the calculation of the {\em
  microscopic} electric field, and from it, nonlinear processes such
as interface induced second harmonic generation
\cite{mezaSecondharmonicGenerationNanostructured2019}.
The theory was generalized to the calculation of the macroscopic
response of periodic binary materials in the {\em retarded} regime, in which
the period of the structure is comparable to the wavelength of
light \cite{perez-huertaMacroscopicOpticalResponse2013}. In this work it was shown
that the concept of a homogenized response does make sense even in
the retarded regime, and that the macroscopic wave equation may be
employed to obtain the photonic band structure of the
system. However, to that end, the spatial dispersion or non-locality of the
macroscopic permittivity has to be accounted for. Nonlocality has been
shown to be important for the description of the response of
nanometric metallic
components \cite{paredes-juarezNonlocalEffectsElectrodynamics2010}. Furthermore,
non-locality of the effective response may be
useful for the design of novel elements such as broadband, omnidirectiona impedance
matching layers
\cite{imUniversalImpedanceMatching2018}.
From the non-local
dependance of the macroscopic response on the wave-vector {$\bm k$},
beyond its dependence on the frequency {$\omega$}, the magnetic
properties of metamaterials made of non-magnetic materials may be
extracted \cite{juarez-reyesMagneticResponseMetamaterials2018} in accordance
to the Landau-Lifshitz approach
\cite{agranovichElectrodynamicsMetamaterialsLandau2009}.

A limitation of the homogenization procedure above is that it is
constrained to binary metamaterials, made of only two components. The
reason is that the geometry of the system is described by a
characteristic function $B(\bm r)$ that takes only two values, 0 and 1, depending
on the material that occupies the position $\bm r$. The theory may
readily be generalized to metamaterials with more components by
renouncing to the factorization into geometry and dynamics, by writing
the theory in terms of the position dependent {\em microscopic}
response $\epsilon(\bm r)$. However, the need for a Hermitian operator
analogous to a Hamiltonian in order to apply Haydock's formalism would
restrict the formalism to media free of dissipation. In
Ref. \cite{mochanRecursiveCalculationOptical2020} it was noted that
using an {\em Euclidean} internal product of {\em wavefunctions}
instead of the usual {\em Hermitian} product, the relevant operators
such as the dielectric response and its longitudinal proyection become
{\em symmetric} in an extended space, in which counterpropagating
waves, with wavevectors $\pm \bm k$ are considered simultaneously. As
symmetric operators share many of their properties with Hermitian
operators, it was possible to reformulate Haydock's formalism for the
calculation of the macroscopic response of multicomponent
metamaterials of arbitrary composition and with an arbitrary number of
components. Nevertheless, that formulation was restricted to the
non-retarded regime.

The purpose of the present paper is to formulate the calculation of
the macroscopic response of multicomponent metamaterials of arbitrary
composition by applyingd the ideas developed in
Ref. \cite{mochanRecursiveCalculationOptical2020} to the retarded
formulation developed in
Ref. \cite{perez-huertaMacroscopicOpticalResponse2013}.

The structure of the paper is the following:
In Sec. \ref{s:theory} we develop the theory. In order to test the
theory, in Sec. \ref{s:appl} we apply it to a simple system, a Bouligand
structure \cite{Neville(1969),Bouligand(1972)}, such as that present
in the cuticles of several arthropods, and consisting of chitin
micro-filaments that are arranged in layers within which the filaments
are aligned, but rotated with respect to the filaments in the layers
above and below, making an helicoidal structure analogous to that
within cholesteric liquid crystals. The system is interesting as it is
a naturaly ocurring photonic crystal whose optical properties include
iridescent metalic-like reflections that furthermore, produce
circularly polarized light.
To perform the calculation we have extended the
computational package {\em Photonic} \cite{Mochan(2016)} to incorporate
the results of Sec. \ref{s:theory}, and we used it to obtain the
macroscopic non-local response,  analyze its structure and obtain
the dispersion relation of the normal modes.
In order test our numerical results, in Sec. \ref{s:annal} we develop
an exact analytical theory to obtain
the macroscopic dielectric tensor and the normal modes of the system,
and we explain their features such as optical activity and
photonic band structure. Finally,
Sec. \ref{s:concl} is devoted to conclusions.

\section{Theory} \label{s:theory}

Following Ref. \cite{perez-huertaMacroscopicOpticalResponse2013}, from
Maxwell's equations we find an integral equation for the electric
field $\bm E$ within the metamaterial, which we write formally as
\begin{equation}
  \label{eq:wave}
  \bm E=-\hat{\mathcal W}^{-1}\frac{4\pi i}{qc\epsilon_h}\bm j^{\ext},
\end{equation}
where $\bm j^{\ext}$ is an external monochromatic current source
oscillating in time with frequency $\omega$ and in space with
some given wavevector $\bm k$, $q=\omega/c$ is the free-space
wave-number corresponding to $\omega$, $\epsilon_h$ is an arbitrary convenient reference
position independent permittivity, and
$\hat {\mathcal W}$ is the {\em wave operator}
\begin{equation}
  \label{eq:waveop}
  \hat{\mathcal W_{ij}}=\frac{\hat{\epsilon}^{ij}}{\epsilon_h}-
  \frac{\partial_i\partial_j-\delta_{ij}\nabla^2}{\epsilon_h q^2},
\end{equation}
with $\hat{\bm\epsilon}$ the {\em microscopic} dielectric operator
corresponding to the position dependent local permittivity $\bm\epsilon(\bm
r)$, which may be a scalar or a tensor with real or complex frequency
dependent components, $\partial_i$ denotes the partial derivative with
respect to the spatial coordinate $r_i$ ($i=$1,2,3 or $x,y,z$), and
$\delta_{ij}$ is Kronecker's delta. As the source $\bm j^{\ext}$ is unrelated
to the induced currents due to the response $\hat{\bm \epsilon}$, it
has no {\em spatial fluctuations}
\cite{mochanElectromagneticResponseSystems1985} related to the texture
of the metamaterial, so it is equal to its average
\begin{equation}
  \label{eq:jm}
  \bm j_M^{\ext}=\bm j_a^{\ext}=\hat {\mathcal P}_a \bm j,
\end{equation}
where we denote macroscopic quantities by the index $M$ and we
introduce the {\em average} operator $\hat{\mathcal
  P}_a$, which together with the {\em fluctuation}
operator ${\mathcal P}_f=\hat 1-\hat {\mathcal P}_a$ forms a couple of complementary
idempotent projectors, $\hat{\mathcal P}_a\hat{\mathcal P}_a=\hat{\mathcal
  P}_a$, $\hat{\mathcal P}_f\hat{\mathcal P}_f=\hat{\mathcal P}_f$,
$\hat{\mathcal P}_a\hat{\mathcal P}_f=\hat{\mathcal P}_f\hat{\mathcal
  P}_a=0$.
Thus, averaging Eq. \eqref{eq:wave} we obtain
\begin{equation}
  \label{eq:waveM}
  \bm E_M=-\hat{\mathcal W}_M^{-1}\frac{4\pi i}{qc\epsilon_h}\bm j^{\ext},
\end{equation}
where we identify the macroscopic inverse wave operator
\begin{equation}
  \label{eq:waveopM}
  \hat{\mathcal W}_M^{-1}=  \hat{\mathcal P}_a\hat{\mathcal W}^{-1}\hat{\mathcal P}_a,
\end{equation}
that is, the inverse macroscopic wave operator is the average of the
inverse of the microscopic wave operator restricted to act on averaged
sources. From the macroscopic wave
operator one can extract the macroscopic dielectric tensor and employ
it to calculate the optical properties of the system.

In Ref. \cite{perez-huertaMacroscopicOpticalResponse2013} the
microscopic wave operator of a binary system was written in terms of
a Hermitian
operator which was built from the characteristic function that defined
the geometry of the system, and its use allowed the use of Haydock's
recursion to obtain a tri-diagonal representation of $\hat{\mathcal
  W}$, from which the macroscopic response was readily
obtained. However, for multicomponent systems that procedure may not
be directly applied, and the presence of dissipation disallows to use
directly the non-Hermitian wave operator within Haydock's scheme. Our
purpose is to obtain a procedure that may be applied to multicomponent
systems of arbitrary geometry and composition. Thus, it must be
capable to deal with dissipation.
In Ref. \cite{mochanRecursiveCalculationOptical2020} a similar problem
was solved by redefining the inner product of two fields. Instead of
the usual Hermitian product
\begin{equation}
  \label{eq:hermitian}
  \braket{\psi|\phi}=\frac{1}{V}\int d^3r\, \psi^*(\bm r)\phi(\bm r) \quad\text{(Hermitian)}
\end{equation}
between two {\em states} $\ket{\psi}$ and $\ket{\phi}$ corresponding
to the scalar wavefunctions $\psi(\bm r)$ and $\psi(\bm r)$ within a
volume $V$, we introduce the {\em
  Euclidean} product
\begin{equation}
  \label{eq:euclidean}
  \braket{\psi|\phi}=\frac{1}{V}\int d^3r\, \psi(\bm r)\phi(\bm r),\quad\text{(Euclidean)}
\end{equation}
notwithstanding the fact that for complex valued fields it is not positive definite and it
may produces complex values for the product of a state with
itself. The generalization to vector valued states is
immediate.  The motivation for using this product is that the matrix elements of the wave
operator between any two vector valued states $\ket{\bm\psi}$ and
$\ket{\bm\phi}$ represented in real space by the functions $\psi_i(\bm
r)$ and $\phi_i(\bm r)$, and in reciprocal space by the Fourier
transforms $\psi_i(\bm p)$ and $\phi_i(\bm p)$, where $\bm p$ is the
wavevector, is
\begin{equation}
  \label{eq:psiWphi}
  \braket{\bm\psi|\hat{\mathcal W}|\bm\phi}=\int d^3p'\int d^3p\, \psi_i(\bm p')
  \left(\frac{\epsilon^{ij}(-\bm p'-\bm p)}{\epsilon_h}-\delta(\bm p'+\bm
      p)\frac{p^2\delta_{ij}-p_ip_j}{\epsilon_h q^2}\right)\phi_j(\bm p),
\end{equation}
where we apply Einstein implicit sumation convention to the repeated
cartesian indices $i$ and $j$ (disregarding if they are subscripts or
superscripts), $\epsilon^{ij}(-\bm p'-\bm p)$ is the
Fourier transform of $\epsilon^{ij}(\bm r)$ with wavevector $\bm
p'-\bm p$ and $\delta(\ldots)$ is Dirac's
delta. Notice that the interchange
$\ket{\bm\psi}\leftrightarrow\ket{\bm\phi}$ leaves the matrix element
invariant. This means that under this product the wave operator is
{\em symmetric}, i.e., it has symmetric matrix elements between any
pair of states. There are important theorems about real symmetric linear
operators that are analogous to those for Hermitian operators, and
some of them may be carried over to complex symmetrical matrices. This
allows generalizing Haydock's procedure to complex symmetrical
matrices.

To proceed, we assume that the system is periodic in space. Thus,
$\epsilon^{ij}(-\bm p-\bm p')$ is null unless $\bm p+\bm p'$
is a reciprocal vector of the lattice. Writing $\bm
p=\bm k+\bm G\equiv\bm G_{\bm k}$, then the only
relevant value for $\bm p'$ are of the form $\bm p'=-\bm k-\bm G'=-\bm
G'_{\bm k}$, where $\bm k$ is a Bloch's vector
and $\bm G$ and $\bm G'$ are reciprocal vectors, and we may rewrite Eq. \eqref{eq:psiWphi} as
\begin{equation}
  \label{eq:psiWphi1}
  \braket{\bm\psi|\hat{\mathcal W}|\bm\phi}=
\int_{BZ} d^3k \sum_{\bm G'}\sum_{\bm G}
  \psi_{i,-\bm G'_{\bm k}}
  \left(\frac{\epsilon^{ij}_{\bm G'-\bm G}-\bm G_{\bm k}^2(\mathcal
      P^T_{\bm G'\bm G}(\bm k))_{ij}}
    {\epsilon_h q^2}\right)\phi_{j,{\bm G_{\bm k}}},
\end{equation}
where $\bm\epsilon_{\bm G'-\bm G}$ is the coefficient of the Fourier
series of $\bm\epsilon(\bm r)$ corrersponding to the reciprocal
vector $\bm G'-\bm G$,
\begin{equation}
  \label{eq:PT}
  (\mathcal P^T_{\bm G'\bm G}(\bm k))_{ij}=\delta_{\bm G',\bm
    G}\delta_{ij} - (\mathcal P^L_{\bm G\bm G'}(\bm k))_{ij}
\end{equation}
is the transverse projector for a wave with wavevector $\bm G_{\bm
  k}\equiv\bm k+\bm G$, and
\begin{equation}
  \label{eq:PL}
  (\mathcal P^L_{\bm G'\bm G}(\bm k))_{ij}=\delta_{\bm G',\bm
    G} \frac{(\bm G_{\bm k})_i (\bm G_{\bm k})_j}{(\bm k+\bm G)^2}
\end{equation}
is the longitudinal projector.
Notice that in this expressions a Bloch's vector
$\bm k$ is only coupled to $-\bm k$ and viceversa, so we can restrict
ourselves to {\em states} consisting of two counterpropagating Bloch
waves ($\pm$)
with fixed Bloch vectors $\pm \bm k$ and thus eliminate the integral
over the Brillouin zone (BZ) in Eq. \eqref{eq:psiWphi1}. We write such
pair of states employing a spinor like notations in terms of 2-vectors
\begin{equation}
  \label{eq:spinor}
  \ket{\bm\phi}\to
  \begin{pmatrix}
    \bm\phi_{+,\bm G}\\
    \bm\phi_{-,\bm G}\\
  \end{pmatrix},
\end{equation}
each of whose components is a vector field that depends on the
reciprocal vectors $\bm G$,
with corresponding inner product
\begin{equation}
  \label{eq:inner}
  \braket{\bm\psi|\bm\phi}=\sum_{\bm G}
  (\bm\psi_{-, -\bm G}\cdot \bm \phi_{+,\bm G}
  +\bm\psi_{+,-\bm G}\cdot \bm \phi_{-,\bm G}).
\end{equation}
The wave operator in this space may then be be represented as a $2\times 2$ matrix
\begin{equation}
  \label{eq:W2x2}
  \mathcal W_{\bm G\bm G'}=
  \begin{pmatrix}
    \frac{\bm\epsilon_{\bm G-\bm G'}}{\epsilon_h}-\frac{\bm G_{\bm
    k}^2 \mathcal P^T_{\bm G\bm G'}(\bm k)}{\epsilon_h q^2} & 0\\
    0&\frac{\bm\epsilon_{\bm G-\bm G'}}{\epsilon_h}-\frac{\bm G_{-\bm
    k}^2 \mathcal P^T_{\bm G\bm G'}(-\bm k)}{\epsilon_h q^2}
  \end{pmatrix}
\end{equation}

In order to tame the growth of $\mathcal W$ with $\bm G$ it is
convenient to rewrite \eqref{eq:W2x2} as
\begin{equation}
  \label{eq:gHg}
  \hat{\mathcal W}= (\hat {\bm 1}-\hat {\bm H}\hat {\bm g})\hat {\bm g}^{-1},
\end{equation}
where the inverse of the operator $\hat{\bm g}$ is
\begin{equation}
  \label{eq:g1}
  {\bm g}^{-1}_{\bm G'\bm G}=
  \begin{pmatrix}
    \bm 1-\frac{G_{\bm k}^2}{\epsilon_h q^2}\mathcal P^T_{\bm G\bm G}(\bm
    k) & 0\\
    0&\bm 1-\frac{G_{-\bm k}^2}{\epsilon_h q^2}\mathcal P^T_{\bm G\bm G}(-\bm
    k),\\
  \end{pmatrix}\delta_{\bm G'\bm G}
\end{equation}
where $\bf 1$ is the cartesian unit matrix, while
\begin{equation}
  \label{eq:H}
  \hat{\bm H}=\frac{\epsilon_h\hat{\bm 1}-\hat{\bm\epsilon}}{\epsilon_h}.
\end{equation}

Notice that $\hat {\bm H}$ and $\hat{\bm g}$ correspond to symmetric operators, but
not so their product $\hat{\bm H}\hat {\bm g}$. Nevertheless, introducing yet
another inner product,
\begin{equation}
  \label{eq:inner2}
  (\bm\psi|\bm\phi)=\braket{\bm\psi|\hat {\bm g}|\bm\phi}
\end{equation}
interpeting $\hat {\bm g}$ as a {\em metric}, the operator $\hat{\bm
  H}\hat {\bm g}$
becomes symmetric, i.e., $(\bm \psi|\hat {\bm H}\hat {\bm g}|\bm
\phi)=\braket{\bm \psi|\hat
  {\bm g}\hat {\bm H}\hat {\bm g}|\bm \phi}=\braket{\bm \phi|\hat
  {\bm g}\hat {\bm H}\hat {\bm g}|\bm \psi}=(\bm \phi|\hat {\bm H}\hat
{\bm g}|\bm \psi)$ for any pair of spinor
states $\ket{\bm \psi}$ and $\ket{\bm \phi}$. The metric $\hat
{\bm g}$ is readily obtained from Eq. \eqref{eq:g1}, and it is a diagonal
matrix in reciprocal space with {\em elements}
\begin{equation}
  \label{eq:g}
  \bm g_{\bm G'\bm G}=
  \begin{pmatrix}
    \frac{\epsilon_h q^2 \bm 1-G_{\bm k}^2 {\mathcal P}^L_{\bm G\bm G}(\bm k)}{
    \epsilon_h q^2 -G_{\bm k}^2}&0\\
    0&\frac{\epsilon_h q^2 \bm 1-G_{-\bm k}^2 {\mathcal P}^L_{\bm G\bm G}(-\bm k)}{
    \epsilon_h q^2 -G_{-\bm k}^2}
  \end{pmatrix}\delta_{\bm G'\bm G},.
\end{equation}

To calculate the inverse macroscopic wave operator,
\begin{equation}
  \label{eq:WM1}
  \hat{\mathcal W}_M^{-1}=\hat {\mathcal P}_a\hat{\bm g}(\hat {\bm
    1}-\hat {\bm H}\hat {\bm g})^{-1}\hat {\mathcal P}_a,
\end{equation}
we start by defining a normalized
{\em macroscopic} starting state $\ket{\bm \phi_0}$, a spinor consisting of two
counterpropagating plane waves with wavevectors $\pm\bm k$, with corresponding
polarizations $\hat{\bm e}_\pm$ and
normalized so that
\begin{equation}
  \label{eq:norm}
  \braket{\bm \phi_0|\bm \phi_0}=1.
\end{equation}
Then, we start Haydock's iteration with the same state but normalized
according to the metric $\hat {\bm g}$,
\begin{equation}
  \label{eq:0}
  \ket{0}=\frac{1}{b_0} \ket{\bm\phi_0},
\end{equation}
where
\begin{equation}
  \label{eq:b0}
  b_0^2=\braket{\bm\phi_0|\hat {\bm g}|\bm\phi_0}.
\end{equation}
Starting from $\ket{0}$ and defining $\ket{-1}\equiv0$, we define
iteratively the states $\ket{n}$ through the Haydock recursion
\begin{equation}
  \label{eq:haydock}
  b_{n+1}\ket{n+1}=\hat {\bm H}\hat {\bm g}\ket{n}-a_n\ket{n}-b_n\ket{n-1},
\end{equation}
where the coefficients $a_n$ and $b_{n+1}$ are chosen so that the state
$\ket{n+1}$ is orthogonal to all previous states and is itself normalized,
i.e.,
\begin{equation}
  \label{eq:normH}
  \braket{n|\hat {\bm g}|m}=\delta_{nm}.
\end{equation}
Due to the symmetry of $\hat{\bm H}\hat{\bm g}$, the coefficient of
$\ket{n-1}$ in Eq. \eqref{eq:haydock} is simply the coefficient $b_n$
obtained in the previous step, and there is no contribution from
states $\ket{n-2}$, $\ket{n-3}$, etc.
Notice that in this basis the wave operator is represented by a
tridiagonal symmetric matrix
\begin{equation}
  \label{eq:tridiag}
  (\mathcal W_{nm})=
  \begin{pmatrix}
    1-a_0 &-b_1  &0     &0    &\cdots\\
    -b_1   &1-a_1 &-b_2   &0    &\cdots\\
    0     &-b_2   &1-a_2 &-b_3  &\cdots\\
    \vdots&\vdots&\vdots&\vdots&\ddots
  \end{pmatrix}
\end{equation}
though its entries are in general complex numbers.
The {\em projected} inverse wave operator is then given by a continued
fraction,
\begin{equation}
  \label{eq:proj}
  \braket{\phi_0|\mathcal W^{-1}|\phi_0}=b_0^2\braket{0|\mathcal
    W^{-1}|0}
  =\frac{b_0^2}{1-a_0-\frac{b_1^2}{1-a_1-\frac{b_2^2}{1-a_2-\ddots}}}-
\end{equation}

From Eqs. \eqref{eq:spinor}, \eqref{eq:inner}  and \eqref{eq:WM1}, we interpret the
result above as
$\hat{\bm e}_-\cdot (\mathcal W_M^{-1})_{++}\cdot\hat{\bm e}_+
+\hat{\bm e}_+\cdot (\mathcal W_M^{-1})_{--}\cdot\hat{\bm e}_-$, where
$\pm$ denote spinor indices. Thus,
repeating the calculation for different polarization vectors
$\hat {\bm e}_\pm$ such as $\hat{\bm e}_+=\hat{\bm x}$, $\hat{\bm y}$,
$(\hat{\bm x}+\hat{\bm y})/\sqrt2$, $(\hat{\bm x}+i\hat{\bm
  y})/\sqrt2$, etc.,  with $\hat{\bm e}_-=\hat{\bm e}_+^*$,
and using the condition of symmetry under the
simultaneous reversal of Bloch's vector $\bm k\leftrightarrow -\bm k$,
and interchange of cartesian indices
$i\leftrightarrow j$, we may obtain all the components of $\mathcal
W_M^{-1}$.  A simple matrix inversion yields then the macroscopic wave
operator, which we interpret using Eq. \eqref{eq:W2x2} in terms of the macroscopic dielectric
tensor as
\begin{equation}
  \label{eq:WM}
  (\mathcal W_M)_{++}=\frac{\bm \epsilon^M(\bm k)}{\epsilon_h}-\frac{k^2\mathcal P^T_{\bm
    0\bm 0}(\bm k)}{\epsilon_h q^2}.
\end{equation}

The procedure above and some others have been incorporated into a free, open access
computational package called {\em Photonic}
written in the {\em Perl} language and using the {\em Perl Data
  Language} number-crunching extension and is
available in a Github repository\footnote{The
  corresponding code can be found in a set of modules named {\tt
  Photonic::WE::S::Haydock}, {\tt ...::Metric}, {\tt ...::GreenP},
{\tt ...::Green}, and {\tt ...::Field} for the case of components with
isotropic permittivities, and the same modules but replacing {\tt
  ::S::} by {\tt ::ST::} for the case on anisotropic components.} \cite{Mochan(2016)}.

\section {Application} \label{s:appl}

\begin{figure}
  \centering
  \includegraphics[width=0.6\textwidth]{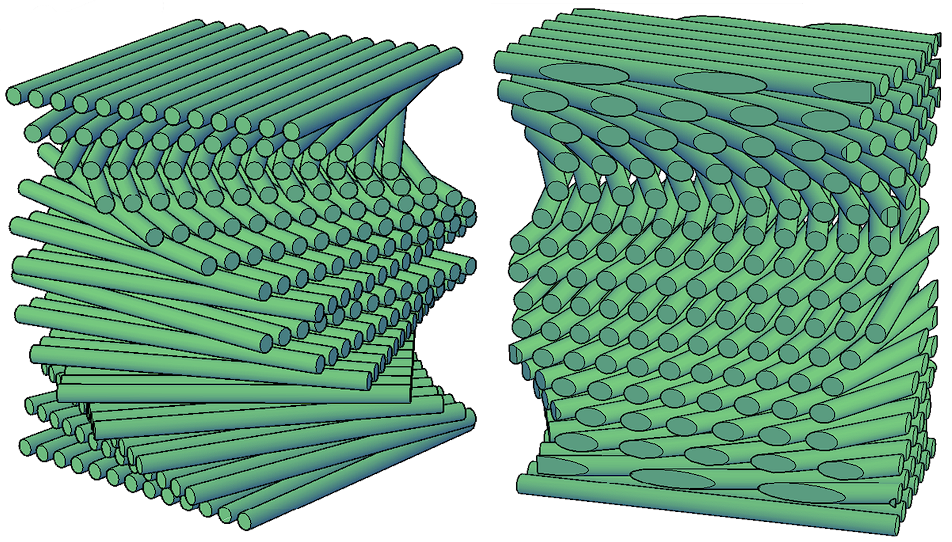}
  \caption{Bouligand structure. Films made of aligned fibers are
    stacked one on top of the other after a relative rotation by a
    constant angle. The left (right) side illustrates a left (right)
    helical structure. (Illustration by Andrea López-Reyna.)}
  \label{fig:schematic}
\end{figure}

To test the theory developed in the previous section and its numerical
implementation we use a simple multicomponent system given by a
Bouligand structure. This kind of structures is present in the cuticle
of several arthropods \cite{Neville(1969),Bouligand(1972)} and
sometimes it is responsible for strong metallic-like reflections which
are frequently iridescent and which furthermore are circularly
polarized. The structures consist of stacked layers, each of which
consists of fibers of chitin, a protein, aligned along the layer and
giving rise to an anisotropic thin film. Subsequent films are stacked
one over the other rotated among themselves around the stacking axis
giving rise to a helical structure as illustrated in
Fig. \ref{fig:schematic}, similar to that of cholesteric
liquid crystals.
Recently, the optical properties of
individual layers was investigated using a homogenization procedure,
and those of the full structure using a transfer matrix approach
\cite{rodriguezAnisotropicOpticalResponse2023}. In this section we
will apply the theory developed in the previous section to obtain the
macroscopic response of Bouligand-like structures.

Notice that the system is periodic as it repeats itself after half a
rotation. Thus, the modes of the system are Bloch waves. For
simplicity, we will consider only Bloch waves moving along the
stacking direction $\hat{\bm s}$. We assume that
each layer has a well defined local dielectric response, which would
be anisotropic. The principal axes of the layer's
response would naturally be the common axis of the fibers $\hat
{\bm f}$, the normal  $\hat{\bm n}=\hat{\bm s}\times\hat{\bm f}$ to
the both fibers and to the stacking direction, and $\hat{\bm s}$,
which we assume coincides with the
$z$ axis. Thus, in these principal axes, the dielectric
tensor would be
\begin{equation}
  \label{eq:layer}
  \bm\epsilon_p=
  \begin{pmatrix}
    \epsilon_{ff}&0&0&\\
    0&\epsilon_{nn}&0\\
    0&0&\epsilon_{zz}
  \end{pmatrix}
  =
  \begin{pmatrix}
    I+A&0&0\\
    0&I-A&0\\
    0&0&\epsilon_{zz}
  \end{pmatrix},
\end{equation}
where we introduced the isotropic $I$ and anisotropic $A$
contributions to the in-plane response.

The dielectric response of a given layer at height $z$ may be obtained
by rotating the response above by an angle $\theta$, which grows with
$z$. Assuming that at $z=0$ $\hat{\bm f}$ points along $\hat{\bm x}$,
we write $\theta(z)=G_0z$ where $G_0=\pm 2\pi/a$ with $a$ the pitch
of the helix, i.e., the distance required for a full turn, and the
sign of $G_0$ is positive (negative) for right (left) helices. Thus, the
actual dielectric tensor at height $z$ would be
\begin{equation}
  \label{eq:epsz}
  \bm \epsilon(z)=\bm R(z) \bm \epsilon_p\bm R^T(z),
\end{equation}
where
\begin{equation}
  \label{eq:R}
  \bm R(z)=
  \begin{pmatrix}
    \cos G_0z&-\sin G_0z&0\\
    \sin G_0z&\cos G_0z&0\\
    0&0&1
  \end{pmatrix}
\end{equation}
is a rotation matrix. Confining our attention to the in-plane response, the result is
\begin{equation}
  \label{eq:epsz1}
  \bm\epsilon(z)=
  I
  \begin{pmatrix}
    1&0\\
    0&1\\
  \end{pmatrix}
  +
  A
  \begin{pmatrix}
    \cos 2G_0z&\sin 2G_0z\\
    \sin 2G_0z& -\cos 2G_0z
  \end{pmatrix},
\end{equation}
a height independent term propoprtional to the identity and a term
that changes sinusoidally with a
spatial period $a/2$ corresponding to a reciprocal vector $2G_0$.

\begin{figure}
  \centering
  \includegraphics[width=.6\textwidth]{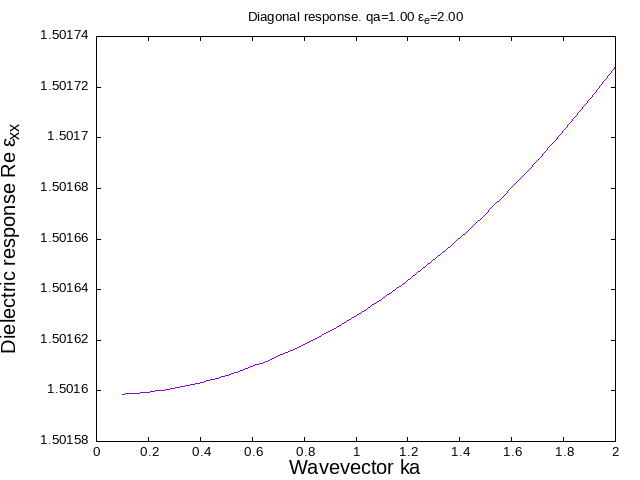}\\
  \includegraphics[width=.6\textwidth]{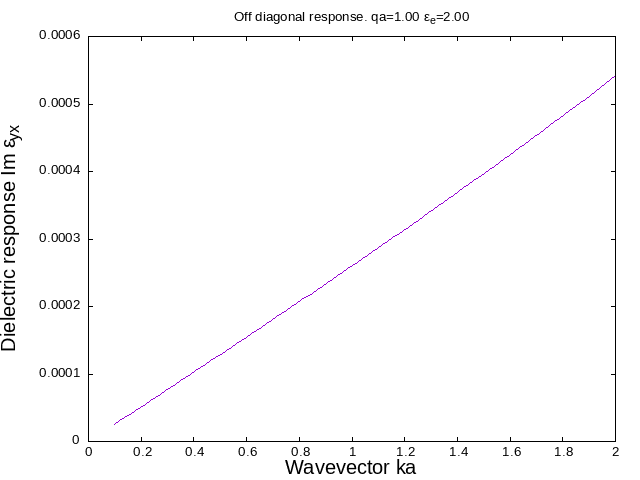}
  \caption{Diagonal components
    $\epsilon^M_{xx}(\omega,\bm k)=\epsilon^M_{yy}(\bm k,\omega)$
    (top) and imaginary part of the offdiagonal components
    $\epsilon^M_{yx}=-\epsilon^M_{xy}$ (bottom)
    of the macroscopic permitivity as
    a function of the wavevector $\bm k=k\hat{\bm z}$ for a system
    with $N=11$ anisotropic layers per revolution with $I=1.5$,
    $A=0.5$, and frequency $\omega=c/a$  }
  \label{fig:epssmall}
\end{figure}

To test the procedure detailed in the preceeding section we performed
a numerical calculation using the {\em Photonic} package \cite{Mochan(2016)},
assuming a unit cell consisting of $N>2$ layers numbered $n=0\ldots N-1$, of equal
width $d$. The $n$-th layer is located at a height $z_n=n d$ and is
rotated by an angle $\theta_n=G_0 n d$, and thus has a dielectric
tensor $\bm\epsilon_n=\bm\epsilon(nd)$ (Eq. \eqref{eq:epsz1}) that differs from that of
other layers due to its different orientation. In the upper panel of
Fig. \ref{fig:epssmall} we show the diagonal
component of the numerically calculatted macroscopic permitivity
$\epsilon^M_{xx}(\omega,\bm k)$ which equals $\epsilon^M_{yy}(\omega,\bm k)$ as a
function of the wavevector $\bm k=k\hat{\bm z}$ for a fixed frequency
$\omega=c/a$ for a right handed Bouligand system with isotropic contribution to the
microscopic permittivity $I=1.5$ and anisotropic contribution
$A=0.5$.
The results are already well converged for $N=11$
and using only 11 pairs of Haydock coefficients but much larger $N$'s
may be used to better approximate a continuosly rotating helix. A non-retarded calculation would
trivially yield an in-plane macroscopic response equal to the average
over all the layers, that is $\bm\epsilon^M=I\hat{\bm 1}$. Our results show
a small deviation at $k=0$ due to the finite frequency and an additional small
quadratic deviation $\propto k^2$ for small $k$. This quadratic dependence on the wavevector
is a non-local effect than may be interpreted in terms of an effective
magnetic response of the spatially dispersive system
\cite{juarez-reyesMagneticResponseMetamaterials2018}. As the system has
no dissipation, its diagonal response is real.

In Fig. \ref{fig:epssmall} we also show the non-diagonal macroscopic response
$\epsilon^M_{yx}$ (lower panel) for the same system as in Fig. \ref{fig:epssmall},
which turns out to imaginary, linear in $k$ for small $k$ and opposite to $\epsilon^M_{xy}$, so
that in the absence of dissipation $\bm \epsilon^M$ is a Hermitian
matrix.

Notice that the fact that the offdiagonal components of the
permittivity tensor turned out to be antisymmetric does not
contradict the statement in the previous section that with the
Euclidean metric the matrix elements of the wave operator and related
operators are symmetric, as the symmetry operation corresponds to an exchange of
Cartesian indices $i\leftrightarrow j$ {\em and} a change of sign of the
wavevector $\bm k\leftrightarrow -\bm k$ as stated in the previous
section. Thus, the symmetric part of $\bm\epsilon^M$ should be an even
function of the wavevector and its antisymmetric part should be an odd
function. The quadratic and linear behaviors shown in
Fig. \ref{fig:epssmall} confirm this
behaviour. Furthermore, in the absence of dissipation, $\bm\epsilon^M$
should be Hermitian. Thus, its symmetric part should be real and its
antisymmetric part should be imaginary.
Off-diagonal imaginary antisymmetric contributions to the response are
an indication of optical activity \cite{LandauXIIS104}, as could
have been expected for Bouligand structures due to their chiral
geometry.

\begin{figure}
  \centering
  \includegraphics[width=.6\textwidth]{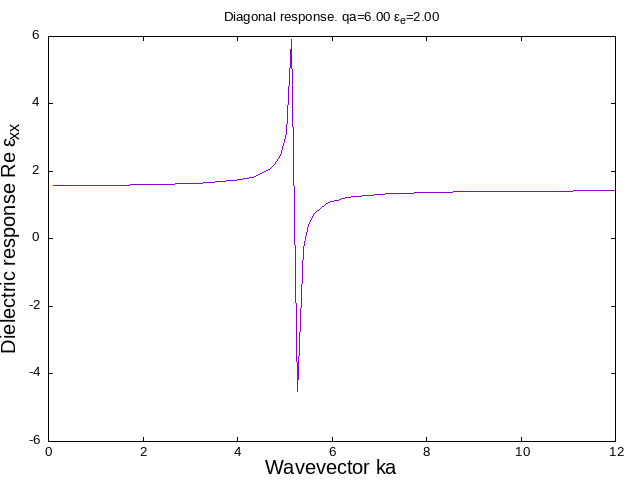}
  \includegraphics[width=.6\textwidth]{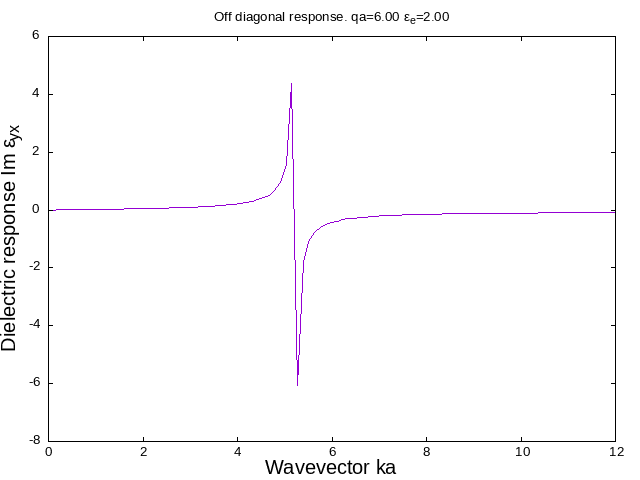}
  \caption{Diagonal (top) and off-diagonal (bottom) macroscopic response for the same system as in
    Fig. \ref{fig:epssmall}, but for a larger frequency $\omega=6c/a$ and larger wavevectors.}
  \label{fig:epslarge}
\end{figure}

The results above correspond to frequencies and wavevectors that are
small with respect to the spatial period of the system. In
Figs. \ref{fig:epslarge} we show the
equivalent results for a larger frequency $\omega=6c/a$ and a larger
wavevector range $0<k<12/a$. Notice that both diagonal and
off-diagonal components have a resonance above $k\approx5/a$.

The normal modes of the structure may be obtained from the macroscopic
wave equation in the absence of sources,
\begin{equation}
  \label{eq:waveEq}
  k^2 \bm E =  q^2 \bm \epsilon^M(\omega,\bm k) \bm E,
\end{equation}
which shows that $k^2/q^2$ should be an eigenvalue of the macroscopic
dielectric tensor $\bm \epsilon^M(\omega,\bm k)$. Given the shape of
the in plane projection $\bm \epsilon^M$, a real multiple of the identity plus an
imaginary antisymmetric matrix, its eigenvalues are of the form
$\text{Re}\,\epsilon_{xx}\pm\text{Im}\epsilon_{yx}$, corresponding to
right circular polarization, with $\bm E\propto (1,i)$, or left
circular polarization, with $\bm E\propto (1,-i)$, as seen from the $z$ axis looking
downward towards the $xy$ plane. Notice that with this convention, a
wave with right (left) circular polarizatin would have positive
(negative) helicity if it propagates in the $z$ direction and a
negative (positive) helicity if it propagates in the $-z$
direction.
\begin{figure}
  \centering
  \includegraphics[width=.5\textwidth]{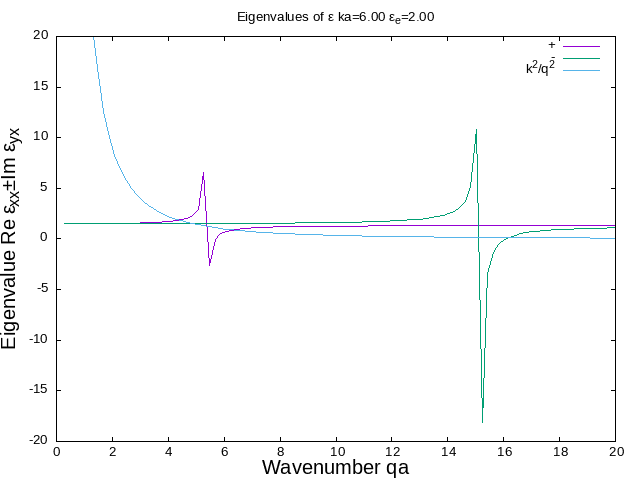}\\
  \includegraphics[width=.5\textwidth]{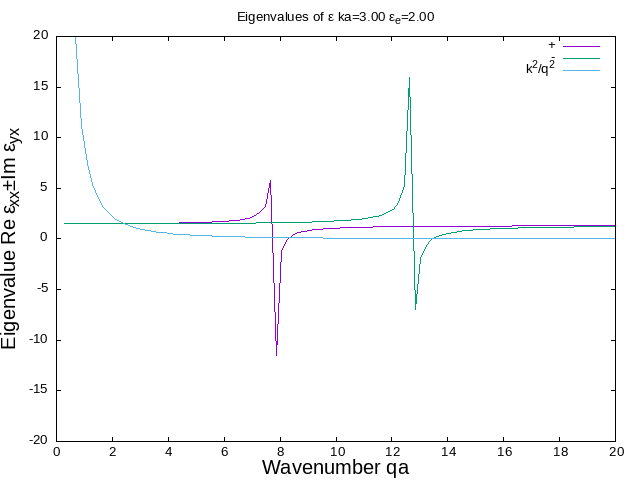}\\
  \includegraphics[width=.5\textwidth]{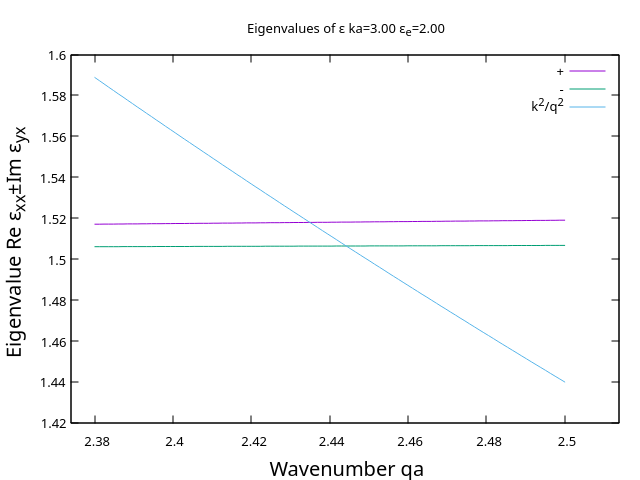}
  \caption{Eigenvalues of $\bm \epsilon^M$ for right ($+$) and left
    ($-$) circular polarizations and for different values
    $k=6/a$ (top panel) and $k=3/a$ (middle and bottom panels) of the wavevector, as
    a function of frequency $\omega=qc$. Also shown are the
    corresponding values of $k^2/q^2$. The right panel is a
    detail of the central panel.}
  \label{fig:eigen}
\end{figure}
The only difficulty for solving the dispersion relation
\eqref{eq:waveEq} is that the eigenvalues
depend on $\bm k$ due to spatial dispersion. In Fig. \ref{fig:eigen}
we show the eigenvalues of $\bm \epsilon^M$ for the two circular
polarizations and for two values of $k$. We also show a plot of
$k^2c^2/\omega^2$. Its intersection with the eigenvalues yields the
frequency $\omega(\bm k)$ for the chosen wavevector $\bm k$ and
polarization. The procedure
could be repeated for several wavevectors to obtain the full
dispersion relation.

\begin{figure}
  \centering
  \includegraphics[width=.5\textwidth]{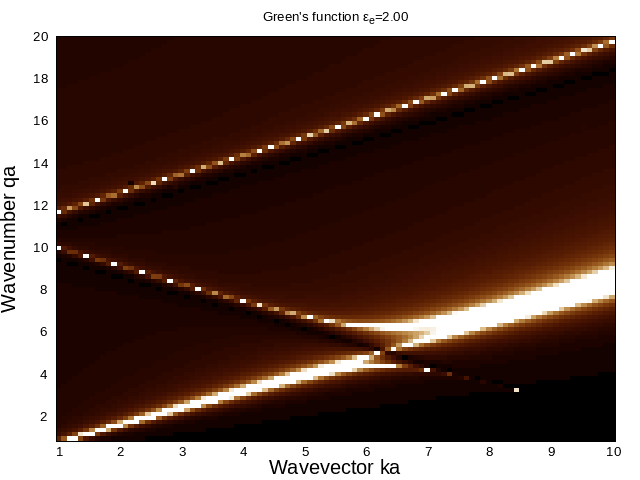}\\
  \includegraphics[width=.5\textwidth]{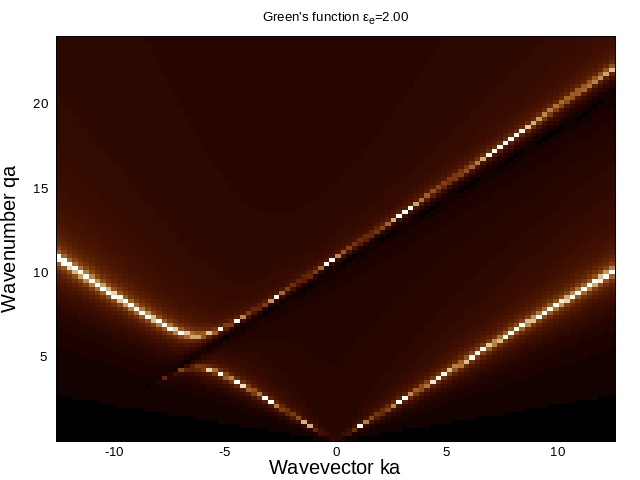}\\
  \includegraphics[width=.5\textwidth]{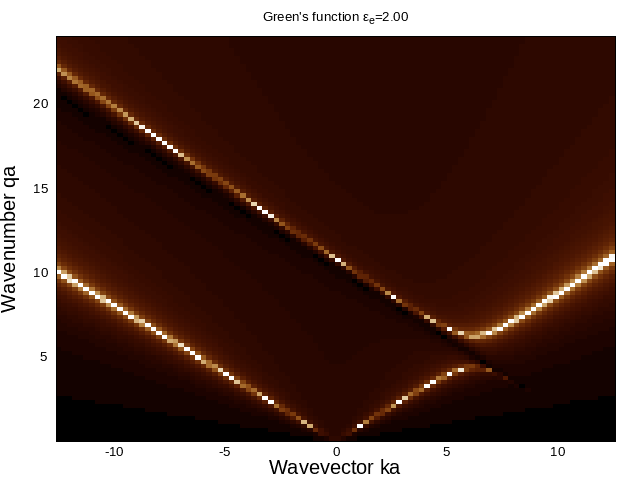}
  \caption{Determinant of the macroscopic inverse wave
    operator $\det \mathcal W_M^{-1}$ for the same
    system as in Fig. \ref{fig:epssmall} as a function of frequency
    $\omega$ and wavevector $\bm k=k\hat{\bm z}$ (top panel). To bound the
    singularities, an imaginary part was added to the
    dielectric function. $\det \mathcal
    W_M^{-1}$ projected into right and left circular polarized states
    (middle and bottom panels).}
  \label{fig:green}
\end{figure}

Instead of searching for the intersections above, we can visualize the
normal modes by plotting the determinant of the Green's
function corresponding to the macroscopic wave operator, $\det \mathcal W_M^{-1}(\omega,k)$,
as the modes correspond to the singularities of the wave operator. For
visualization purposes, we bound the poles of the Green's function by
introducing some dissipation, adding an imaginary part to the
diagonal part of the microscopic response. This illustrates the
fact that our procedure is able indeed to deal with dissipative
multicomponent systems. The results are shown in the top panel of
Fig. \ref{fig:green}.
A rich photonic band structure corresponding to the singularities of
$\mathcal W_M$ may be observed. Its interpretation becomes simpler by
projecting the Green's function into right and left circular
polarizations. Notice that there
are photonic gaps at $k=\pm G_0$, the boundary of the 1D Brillouing
zone, for which waves with negative helicity may not propagate within
the structure. Accordingly, we expect a strong reflectance in a finite
Bouligand structure at the corresponding frequencies. It is
interesting to note that in this system there is negligible contrast
between the response of adjacent layers; they have the same
composition and they only differ by a small rotation. Nevertheless,
the system is periodic and therefore it develops a photonic structure
with allowed and forbidden bands. Furthermore, the dispersion relation
for a given rotation sense is not symmetric under reversal of the
propagation direction (for a given helicity it is symmetric).

\section {Analytical results} \label{s:annal}

In the previous section we obtained the macroscopic response and the
normal modes of a Bouligand structure by using a general purpose
theory developed for structures with an arbitrary number of different
components of arbitrary geometry and composition, and employing its
numerical implementation. Nevertheless, to test the validity of the
results, it is important to compare them with those of alternative
approaches. In this section we develop an analytical solution of the
simple structure studied in section \ref{s:appl}  and compare the
results of both approaches.

\begin{figure}
  \centering
  \includegraphics[width=.6\textwidth]{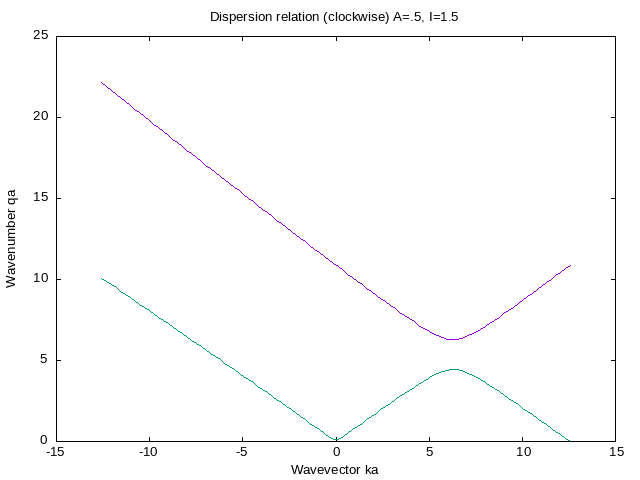}
  \caption{Exact dispersion relation for a left polarized wave
    propagating with wavevector $\bm k$ along the $z$
    direction within the same system as in Fig. \ref{fig:epssmall}.}
  \label{fig:exactleft}
\end{figure}

Consider a circularly polarized plane wave propagating along
$z$ with in-plane field
\begin{equation}
  \label{eq:Eext}
\bm E_{k\sigma}(z)=\frac{E_{k\sigma}}{\sqrt 2}
\begin{pmatrix}
  1\\\sigma i
\end{pmatrix}
e^{ikz},
\end{equation}
where $\sigma=1$ (-1) for right (left) circular
polarization.
Its interaction with a continuously rotating Bouligand structure
described by the response given by Eq. \eqref{eq:epsz1} yields
\begin{equation}
  \label{eq:PmG}
    \bm \epsilon \bm E_{k\sigma}(z)
    =
      I\bm E_{k\sigma}e^{ikz}
      +
      A\frac{E_{k\sigma}}{\sqrt 2}
      \begin{pmatrix}
        1\\-\sigma i
      \end{pmatrix}
      e^{i(k+\sigma 2G_0)z}.
\end{equation}
Curiously, a right polarized wave with wavevector $\bm k$ is single-scattered into a left polarized
wave with wavevector $\bm k+2\bm G_0$, while a left polarized wave is
scattered into a right polarized wave wave with wavevector $\bm k-2\bm
G_0$. Thus, even allowing multiple scattering, there would only be two coupled waves
propagating within the system with opposite polarizations and with
wavevectors that differ by $\pm 2G_0$. The homogeneous wave equation
\begin{equation}
  \label{eq:wave2}
  (\partial_z^2+q^2\bm\epsilon(z))\bm E=0,
\end{equation}
can thus be written as an exact $2\times2$ matricial system. For
example, consider a left polarized wave propagating along $z$ with
wavevector $\bm k$ and amplitude $E_{0-}$. It would only be
coupled within a right Bouligand structure to a right polarized wave with
wavevector $\bm k-2\bm G_0$ and amplitude, say $E_{s+}$. Thus, Eq. \eqref{eq:wave2} becomes
\begin{equation}
  \label{eq:acop}
  \begin{pmatrix}
    -k^2+q^2 I&q^2 A\\
    q^2 A& -(k-2G_0)^2+q^2 I
  \end{pmatrix}
  \begin{pmatrix}
    E_{0-}\\E_{s+}
  \end{pmatrix}=0,
\end{equation}
which yields immediately the dispersion relation
\begin{equation}
  \label{eq:disp}
  q^4(I^2- A^2)-q^2 I(k^2+(k-2G_0)^2+k^2(k-2G_0)^2=0,
\end{equation}
a quadratic equation in $q^2$ which yields two bands for each
 $k$ with frequencies $\omega_\pm=q_\pm,c$ where
\begin{equation}
  \label{eq:sol}
  q^2_\pm=\frac{I(\tilde k^2+G_0^2)\pm\sqrt{(\tilde
      k^2+G_0^2)^2 I^2 - (I^2-A^2) (\tilde
      k^2-G_0^2)^2}} {I^2-A^2}.
\end{equation}
Here we introduced a translated wavevector  $\tilde k\equiv k-G_0$ to
simplify the expression. In Fig. \ref{fig:exactleft} we show the exact
dispersion relation for this case. A comparison with the corresponding
numerical result, the bottom panel of Fig. \ref{fig:green} shows an
almost perfect agreement.

From Eq. \eqref{eq:disp} we may easily find the bandgap edges at $k=G_0$,
\begin{equation}
  \label{eq:bandedge}
  q_\pm=\frac{G_0}{\sqrt{I\mp A}}.
\end{equation}
The reason is that for $\bm k=\bm G_0$, the scattered field has wavevector
$\bm k-2\bm G_0=-\bm G_0$, and, as usual at the border of the Brillouin zone, the
field is a stationary wave formed by either an even or an odd combination of the
counter propagating components. In the former case, $E_{0-}=E_{s+}$ and the resulting
microscopic field is helical, linearly polarized for all $z$ along the local
direction of the fibers, so that the
response to the field is $I+A$ everywhere, according
to Eq. \eqref{eq:layer}. Thus, $k^2=G_0^2=q_+^2(I+A)$.
Similarly, for the odd combination, the field is linearly polarized
perpendicular to the direction of the fibers, so the appropriate
response is $I-A$ and $G_0^2=q_-^2(I-A)$.

\begin{figure}
  \centering
  \includegraphics[width=.6\textwidth]{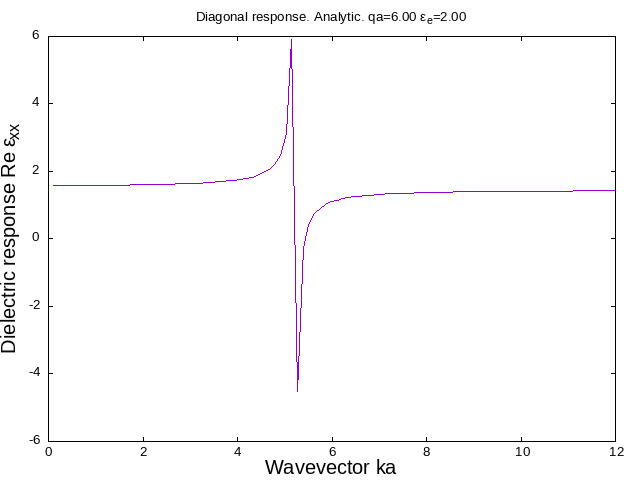}\\
  \includegraphics[width=.6\textwidth]{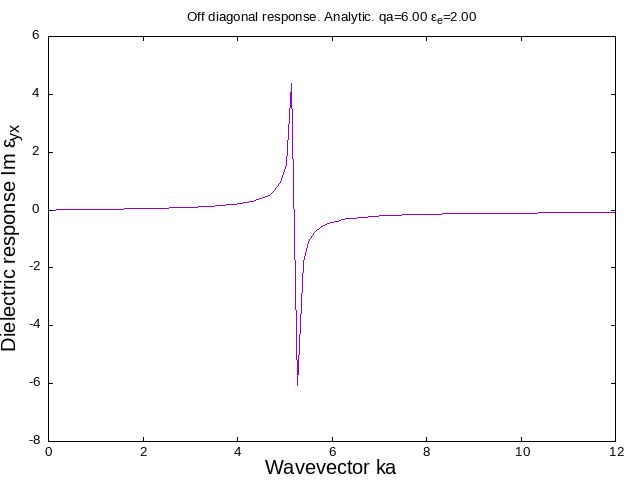}
  \caption{Exact real part of the diagonal diagonal (top) and
    imaginary part of the off-diagonal (bottom) macroscopic dielectric
    response, $\epsilon^M_{xx}=\epsilon^M_{yy}$ and
    $\epsilon^M_{yx}=-\epsilon^M_{xy}$ respectively, of the same
    system as in Fig. \ref{fig:epssmall}, with parameters corresponding to
    Fig. \ref{fig:epslarge}. }
  \label{fig:epsexact}
\end{figure}

Besides the dispersion relation, we may also obtain analytical
expressions for the macroscopic permittivity. To that end, consider an
external current that propagates with wavevector $\bm k$ along the
$z$ direction and with right polarization, i.e.
\begin{equation}
  \label{eq:jex}
  \bm j_{\ext}=\frac{j_+}{\sqrt2}
  \begin{pmatrix}
    1\\i
  \end{pmatrix}
  e^{ikz}.
\end{equation}
This current induces a right polarized wave with wavevector
$\bm k$ which is scattered by the Bouligand structure into a left
circularly polarized wave with wavevector $\bm k+2\bm G_0$. Calling
the respective amplitudes $E_+$ and $E_-$, we rewrite the
inhomogeneous wave equation
\begin{equation}
  \label{eq:inhomwave}
  \partial_z^2\bm E+q^2\bm\epsilon(z)\bm E=-\frac{4\pi i q}{c}\bm j_\ext
\end{equation}
as
\begin{equation}
  \label{eq:inhomwave2}
  \begin{pmatrix}
    -k^2+q^2 I&q^2 A\\
    q^2 A&-(k+2G_0)^2+q^2 I
  \end{pmatrix}
  \begin{pmatrix}
    E_+\\E_-
  \end{pmatrix}
  =
  -\frac{4\pi i q}{c}
  \begin{pmatrix}
    j_+\\0
  \end{pmatrix}.
\end{equation}
Eliminating $E_-$ from this system of equation we obtain
\begin{equation}
  \label{eq:E+}
  \left(
    -k^2+q^2I+\frac{q^4 A^2}{(k+2G_0)^2-q^2I}
  \right)
  E_+
  =
  -\frac{4\pi i q}{c} j_+.
\end{equation}
Interpreting this equation as a macroscopic inhomogeneous wave
equation
\begin{equation}
  \label{eq:wave+}
  (-k^2+q^2\epsilon^M_+)E_+=  -\frac{4\pi i q}{c} j_+,
\end{equation}
we may identify the macroscopic permittivity for right circular
polarization
\begin{equation}
  \label{eq:eps+}
  \epsilon^M_+=I+\frac{q^2 A^2}{(k+2G_0)^2-q^2I}.
\end{equation}
Similarly, for left circular polarization we obtain
\begin{equation}
  \label{eq:eps-}
  \epsilon^M_-=I+\frac{q^2 A^2}{(k-2G_0)^2-q^2I}.
\end{equation}
From these, we may readily obtain the permittivity in the cartesian
basis,
\begin{equation}
  \label{eq:epsM}
  \bm\epsilon^M=\frac{1}{2}\left(
    \begin{pmatrix}
      1\\i
    \end{pmatrix}
    \epsilon^M_+
    \begin{pmatrix}
      1,&i
    \end{pmatrix}
    +
    \begin{pmatrix}
      1\\-i
    \end{pmatrix}
    \epsilon^M_-
    \begin{pmatrix}
      1,&-i
    \end{pmatrix}
    \right).
\end{equation}
After some algebra, we obtain analytical expressions for the
macroscopic dielectric function of the Bouligand structure,
\begin{equation}
  \begin{split}
  \label{eq:epsexact}
  \epsilon^M_{xx}&=\epsilon^M_{yy}=
                   I+q^2 A^2\frac{k^2+4G_0^2-q^2I}
                   {(k^2-4G_0^2)^2-2 q^2 I(k^2+4G_0^2)+q^4I^2}\\
    \epsilon^M_{xy}&=-\epsilon^M_{yx}=
                     \frac{4 i k G_0 q^2 A^2}
                     {(k^2-4G_0^2)^2-2 q^2 I(k^2+4G_0^2)+q^4I^2}.
  \end{split}
\end{equation}
We may verify that substitution of this response into the wave
equation yields the same dispersion relation as obtained previously,
Eq. \eqref{eq:sol}. In Fig. \ref{fig:epsexact} we show the exact non-local
macroscopic response corresponding to the same parameters as in Fig. \ref{fig:epslarge}.
The figures are almost indistinguishable. Furthermore, from the
denominators in Eqs. \eqref{eq:epsexact} we can locate the position of
the poles,
\begin{equation}
  \label{eq:kpolo}
  k^2_\infty=q^2 I + 4 G^2 \pm 4 q G\sqrt{I}.
\end{equation}
For the parameters used for Fig. \ref{fig:epslarge}, there is a pole
around $ka=5.22$, as seen in Figs. \ref{fig:epslarge} and
\ref{fig:epsexact}. There is a second pole
around $ka=19.91$ which we have verified is also present in both the numerical
as well as the analytic calculation.
\section {Conclusions} \label{s:concl}
We have been able to extend an efficient homogenization procedure
based on Haydock's recursive algorithm for the calculation of the
macroscopic response of a periodic metamaterial made of an arbitrary
number of components with an arbitrary composition and an arbitrary
geometry. Our extension is based on the observation that using an
Euclidean, rather than a Hermitian inner product to define matrix
elements of operators, all relevant operators become symmetric, even
in the presence of dissipation. However, the inner product couples
Bloch waves that propagate in opposite directions, so our formalism
requires the simultaneous solution of the field equations for
spinor-like fields with two components corresponding to
counter-propagating Bloch waves. A previous work
\cite{mochanRecursiveCalculationOptical2020} followed similar ideas,
but was restricted to the non-retarded regime, while the current work
is applicable to waves of arbitrary frequency and wavevector. The
formalism has been coded into a freely available open source
computational package \cite{Mochan(2016)}. As a test of our formalism,
we have applied it to the calculation of the macroscopic response and
the dispersion relation of light within Bouligand-like structures,
such as those found in the cuticle of some arthropods. These
structures consist of anisotropic layers whose principal axes are
continuously rotated as they are stacked on top of each other,
yielding chiral helical periodic structures. We found that this
structures display optical activity, and that for one helicity
they have a photonic gap, which explains the circular polarization
upon reflection displayed by the cuticle of some insects. To verify
our results we obtained analytical results for these simple
systems. Besides having an excelent agreement with the numerical results,
encouraging the use with confidence of our Haydock scheme to tackle more
complex systems, they allowed us to understand the structure of the
macroscopic response, its singularities, the photonic band structure,
the polarization of the normal modes, and the position and width of the bandgaps.

\acknowledgments
We are grateful to Andrea López Reyna for lending us her illustrations.
WLM acknowledges the support of DGAPA-UNAM under grant No. IN109822.
GPO thanks to SGCyT-UNNE for financial support through grant PI18F008.

\bibliography{references}
%
\end{document}